\documentclass[12pt]{article}
\usepackage{amsmath,float}
\usepackage{graphicx,subfigure,subcaption}
\usepackage{enumerate}
\usepackage{placeins}
\usepackage{natbib, makecell, booktabs}
\usepackage{url} 
\usepackage{color}
\usepackage{amsbsy}
\usepackage{amsmath}
\usepackage{amssymb}
\usepackage{latexsym}
\usepackage{comment}
\usepackage{graphicx}
\usepackage{amsfonts}
\usepackage{url}
\usepackage{bm}
\usepackage{bbm}
\usepackage{multirow}
\usepackage{bm,authblk}
\usepackage{adjustbox}
\usepackage{ragged2e}
\usepackage[symbol]{footmisc}
\usepackage[linesnumbered,ruled,vlined]{algorithm2e}

\newcommand{\blind}{1}

\newcommand{\scsection}[1]{%
  \vspace{1.5em}%
  \noindent{\bfseries\Large #1}%
  \par\vspace{0.75em}%
}

\newtheorem{theorem}{Theorem}
\newtheorem{lemma}{Lemma}

\newtheorem{proposition}{Proposition}
\newtheorem{assumption}{Assumption}

\graphicspath{{./figs/}}
\begin{document}

\def\spacingset#1{\renewcommand{\baselinestretch}%
{#1}\small\normalsize} \spacingset{1}


\if1\blind
{
  \title{\bf Bi-Gaussian Mirrors for False Discovery Rate Control}

 \author{Yujia Wu} 
 \author{Panxu Yuan
 }  
 \author{Binyan Jiang\thanks{Corresponding author: Binyan Jiang. Email: by.jiang@polyu.edu.hk.}}

  \affil{Department of Data Science and Artificial Intelligence \\ The Hong Kong Polytechnic University, Hong Kong}
  \renewcommand*{\Authands}{ and }  
  \date{}  
  \maketitle
} \fi

\if0\blind
{
  \bigskip
  \bigskip
  \bigskip
  \begin{center}
    {\LARGE\bf Bi-Gaussian Mirrors for False Discovery Rate Control}
\end{center}
  \medskip
} \fi

\bigskip
\begin{abstract}

{Effectively controlling the false discovery rate (FDR) in high-dimensional variable selection is a fundamental statistical problem that has garnered significant research interest. In this paper, we propose a novel, user-friendly, and computationally efficient method called Bi-Gaussian Mirrors (BGM), which offers a conceptually simple yet powerful approach for FDR control. Our method makes the first attempt to achieve FDR control in high-dimensional data with complex dependencies, while overcoming key limitations of existing approaches, such as prior knowledge of the joint distribution of data, significant power loss, the need for full symmetry in test statistics, and the theoretical restriction to linear regression models. Additionally, we present a self-guiding procedure designed to enhance the practicality and applicability of the BGM method.
Theoretical guarantees for FDR control and asymptotic power are rigorously established under regularity conditions. Moreover, extensive numerical simulations and two real-data examples demonstrate that the BGM method outperforms existing approaches in terms of finite-sample performance, achieving a superior balance between FDR control and testing power.
}


\end{abstract}

\noindent%

\newpage
\spacingset{1.9} 
\section{Introduction}
\label{sec:intro}

Rapid advances in information technology have led to the widespread emergence of high-dimensional data across various disciplines, including genomics \citep{clarke2008properties,carvalho2008high}, neuroscience \citep{pang2016dimensionality,mackevicius2019unsupervised}, and health care administration \citep{mohammed2010centralized,wang2015outsourcing}, among others.
In such high-dimensional settings, effective statistical inference critically depends on selecting a small subset of the most informative features for predicting the response variable while eliminating redundant ones from the large pool of candidates. This task, known as variable/feature selection, has been widely studied in high-dimensional statistics; see, for example, \cite{Tibshirani1996, Fan2001,Zou2005}, and \cite{Zou2006}, among others.

Let  $ X =(X_1, \cdots, X_p)^{\top}\in\mathbb{R}^p$  be a $p$-dimensional covariate  and $Y\in\mathbb{R}$ be the response variable. To perform variable selection, we focus on testing the following conditional independence hypotheses:
\begin{equation}\label{Conditional}
H_{0j}: Y \perp X_j \mid  X_{-j} \quad \text{v.s.} \quad H_{1j}: Y \not\perp X_j \mid  X_{-j}, \quad \text{for} \quad j =1, \cdots, p.
\end{equation}
Here, $X_j$  
represents the $j$-th component of $ X$, and $ X_{-j}$ denotes the sub-vector of $ X$ after removing $X_j$. 
Under $H_{0j}$, the covariate $X_j$ is unrelated to the response variable $Y$ and thus provides no predictive value for $Y$. We denote $X_j$ as a {\it null} covariate in this case. Conversely, under $H_{1j}$, we refer to $X_j$ as a {\it non-null} covariate. Furthermore, we define $\mathcal{H}_0=\{j: H_{0j}$ is true$\}$ as the set of {\it nulls}  and $\mathcal{H}_1=\{j: H_{1j}$ is true$\}$ as  the set of {\it non-nulls}, respectively.
The aim of variable selection is to recover $\mathcal{H}_1$, which is crucial for reliable statistical inference. As a result, assessing the effectiveness of the selected $\mathcal{H}_1$ has emerged as a key research priority, particularly to achieve robust performance. 
Within this context, adaptive control of the error rate for the selected $\mathcal{H}_1$ has attracted considerable attention, particularly from practitioners in applied fields.

One can observe that, for $j= 1, \cdots, p$, the hypothesis test (\ref{Conditional}) gives rise to a multiple testing problem. When $p$ is large, 
such as when $p/n \to c$ for a positive constant $c$, merely controlling the type I error of each test for $X_j$, as in the classical single hypothesis testing, results in a substantial number of falsely rejected $H_{0j}$. This renders the test results impractical.
To this end, controlling the false discovery rate (FDR) in multiple testing problems has been extensively studied since the seminal work of \cite{Benjamini1995}; see, for example, \cite{storey2002direct, wu2008false, Fan-Han-Gu2012, Fan-Han2017}, and the references therein.
 Let $a \vee b=\max \{a, b\}$ and denote $\#\{\cdot\}$ as the number of elements in a given set. Formally, the FDR is defined as:
\begin{equation}\label{FDR}
\mathrm{FDR}=\mathbb{E}\left[\frac{\#\{j: j \in \widehat{\mathcal{H}}_1 \text{ and } j \in \mathcal{H}_0 \}}{\#\{j: j \in \widehat{\mathcal{H}}_1 \} \vee 1}\right],
\end{equation}
where $\widehat{\mathcal{H}}_1$ denotes the set of  {\it non-null} outcomes obtained from a data-dependent testing or selection procedure. In light of Equation (\ref{FDR}), the primary objective of FDR control is to identify as many {\it non-null} covariates as possible while keeping the expected proportion of falsely rejected $H_{0j}$ at a pre-given level, ensuring the reliability of the selection procedure.

To address this, \cite{Benjamini1995} introduced the Benjamini-Hochberg (BH) method. Let $p_j$ denote the p-value associated with the hypothesis test for $X_j$, and let $p_{(j)}$ represent the p-values $p_j$ arranged in non-decreasing order, with $j=1,\cdots,p$. The BH method identifies a cutoff position $j^*$ such that $p_{(j^*)} \leq (j^*q)/p$ for a predetermined FDR level $q$. It then defines the set $\widehat{\mathcal{H}}_1 = \{j : j \leq j^*\}$, where covariates with smaller p-values are prioritized due to their greater significance in predicting $Y$.
\cite{Benjamini1995} showed that their method can control the FDR at any predetermined level $q$ under the assumption of mutual independence among the $p_j$ values. However, in high-dimensional settings, deriving the null distribution of test statistics is particularly challenging \citep{Sara2014, zhang2014confidence}, and the $X_j$ often exhibit complex dependencies \citep{Fan-Han-Gu2012}, further constraining the applicability of the BH method.

To overcome the limitations of the BH method, two extensively studied classes of approaches have emerged: knockoff-type methods and mirror-type methods. The core idea of these approaches is to assign each covariate $X_j$ ($j=1,\cdots,p$) a variable importance measure, denoted as $\omega_j$,
and then define $W_j = f(\omega_j)$ as the variable importance score, where $f(\cdot)$ is a pre-specified function, to quantify the contribution of $X_j$ to predict $Y$. Intuitively, a larger $W_j$ indicates stronger evidence that $X_j$ significantly influences $Y$. 
Thus, the set of selected variables can be constructed as $\widehat{\mathcal{H}}_1=\{j:W_j>t\}$, and the FDR defined in Equation \eqref{FDR} can be reformulated as:
\begin{equation}\label{FDR-select}
\mathrm{FDR}=\mathbb{E}\left[\frac{\#\{j: W_j>t \text{ and } j \in \mathcal{H}_0 \}}{\#\{j: W_j>t \} \vee 1}\right].
\end{equation}
Over the past decade, this avenue has attracted significant attention. The primary difference between these methods lies in the construction of $W_j$, which is elaborated below.

\underline{\bf The Knockoff-Type Methods}. Since the introduction of the fixed-X knockoff filter by \cite{Barber2015}, knockoff-type methods have been extensively investigated to control the FDR; see, for instance, \cite{Barber2019, 
Sarkar2022, Cao2024SK, fan2025asymptotic} 
and the references therein. Next, we use the fixed-X knockoff filter to illustrate the core idea of the knockoff-type approach. This method achieves FDR control by constructing a set of knockoff variables, denoted as $\widetilde X =(\widetilde X_1, \cdots, \widetilde X_p)^{\top}\in\mathbb{R}^p$. Notably, $\widetilde X$ mimic the correlation structure of the original variables $ X$ while being completely noise variables with respect to the response $Y$. Thus, for each pair of $X_j$ and $\widetilde X_j$, if the impact of $X_j$ on $Y$ is statistically indistinguishable from, or negligible compared to, that of $\widetilde X_j$, the fixed-X knockoff filter identifies $X_j$ as a noise variable, i.e. a {\it null} covariate. 
Consequently, to measure the relative impact of $X_j$ and its knockoff $\widetilde X_j$ on the response $Y$, under Gaussian linear models, \cite{Barber2015} constructed the variable importance score as $W_j = f(\widehat{\beta}_j, \widehat{\beta}_{j+p})$,
where $\widehat{\beta}_j$ and $\widehat{\beta}_{j+p}$ are the estimated regression coefficients for $X_j$ and $\widetilde{X}_j$, respectively, obtained from regressing $[X, \widetilde{X}]$ on $Y$.
Subsequently, \cite{Candes2018} proposed a general model-X knockoff framework to accommodate nonlinear models,
which has been extended to various domains by researchers including \cite{Barber2020robust, Spector2022powerful, Chien2024}, 
among others. Although these knockoff-type methods have demonstrated strong performance, they are hindered by two key limitations: (i) They require prior knowledge of the joint distribution of $X$, which is often impractical to estimate with precision in real-world applications.  (ii) The statistics $W_j$ have to be symmetric around zero under $H_{0j}$, a property that is hard to meet as constructing ``good'' knockoff variables is particularly challenging in real high-dimensional data analysis. 

\underline{\bf The Mirror-Type Methods}. To allow for distribution-free data (addressing limitation (i) of the knockoff-type methods), \cite{Dai2022} and \cite{Xing2023} proposed two mirror-type statistics, which have since been further developed in subsequent studies \citep{Dai2023, harada2024false, wang2024false}. 
Specifically, \cite{Dai2022} proposed a data splitting (DS) mirror method that splits the whole observations $(Y, X)$ into two groups $(Y^{(1)}, X^{(1)})$ and $(Y^{(2)}, X^{(2)})$. Under linear models, they used each group to estimate the regression coefficient for $X_j$, denoted as $\widehat\beta_j^{(1)}$ and $\widehat\beta_j^{(2)}$. Then, they constructed the variable importance score as $W_j=f(\widehat\beta_j^{(1)}, \widehat\beta_j^{(2)})$. This approach was later extended to generalized linear models by \cite{Dai2023}. However, as is well known, splitting data can lead to a loss of power and may result in unstable selection outcomes \citep{wasserman2013all}.
To address these concerns, \cite{Xing2023} developed the Gaussian mirrors (GM) method. For each $X_j$, the GM method constructs two mirror features, $X_j^+=X_j+c_jz_j$ and $X_j^-=X_j-c_jz_j$, where $z_j$ is an independently simulated standard Gaussian random variable, and $c_j$ is a scalar depending on $ X$ and $z_j$. Under linear models, \cite{Xing2023} obtained the estimated regression coefficients for $X_j^+$ and $X_j^-$, denoted as $\widehat\beta_j^+$ and $\widehat\beta_j^-$, respectively, and then constructed $W_j=f(\widehat\beta_j^+,\widehat\beta_j^-)$. 
Although these methods overcome limitation (i) of knockoff-type methods, they still require $W_j$ to be fully symmetric around zero under $H_{0j}$, which substantially complicates the construction of the test statistic in high-dimensional settings. For example, even in linear models, the GM method requires careful design of $c_j$ to ensure the symmetry of $W_j$, as emphasized in Lemma 2 of \cite{Xing2023}.
This challenge becomes more pronounced when extending beyond linear models \citep{Dai2023}.

In this paper, we propose bi-Gaussian mirrors (BGM), a novel and easily implementable method to identify {\it non-null} covariates in high-dimensional data with FDR control, addressing limitations of previous methods. 
The proposed BGM imposes no restriction on $p$ relative to $n$, and is applicable to both linear models and generalized linear models. We model the relationship between covariate $X$ and the response variable $Y$ by $\mathbb{E}(Y) = g( X^\top\beta)$, where $g$ is a pre-specified link function. The key idea of the BGM method is as follows. First, drawing from the GM method, we construct two sets of feature mirrors for each $X_j$, namely $X_{j(1)} = (X_{j(1)}^+, X_{j(1)}^-) = (X_j + \zeta_j^{(1)}, X_j - \zeta_j^{(1)})$ and $X_{j(2)} = (X_{j(2)}^+, X_{j(2)}^-) = (X_j+\zeta_j^{(2)}, X_j-\zeta_j^{(2)})$, where $\zeta_j^{(1)}$ and $\zeta_j^{(2)}$ are independently simulated standard Gaussian random variables.
We then estimate the regression coefficients for $X_{j(1)}$ and $X_{j(2)}$, yielding $\widehat\beta_{j(1)} = (\widehat\beta_{j(1)}^+, \widehat\beta_{j(1)}^-)^\top$ and $\widehat\beta_{j(2)} = (\widehat\beta_{j(2)}^+, \widehat\beta_{j(2)}^-)^\top$, respectively. Next, we calculate two feature Gaussian mirror statistics: $W_j^{(1)} = f(\widehat\beta_{j(1)}^+, \widehat\beta_{j(1)}^-)$ and $W_j^{(2)} = f(\widehat\beta_{j(2)}^+, \widehat\beta_{j(2)}^-)$, where $f(\cdot)$ is a given nonnegative function satisfying $f(x)\geq 0$ for any $x\in\mathcal{R}$. 
The proposed BGM statistic is then defined as $M_j = W_j^{(1)} - \gamma_j W_j^{(2)}$, where $0 < \gamma_j \leq 1$ is a pre-specified weight. 
By replacing $W_j$ with $M_j$ in Equation \eqref{FDR-select}, we can identify {\it non-null} variables while controlling the FDR.
Notably, unlike the GM method, our approach does not incorporate the scalar $c_j$  when constructing the feature Gaussian mirrors 
 $X_{j(1)}$ and $X_{j(2)}$, 
nor does it require any information about the asymptotic distributions of the parameter estimates. This simplification makes our approach more straightforward and potentially adaptable for FDR control in more complex scenarios.
\begin{figure}[H]
\centering
\includegraphics[width=0.75\textwidth,height=5.5cm]{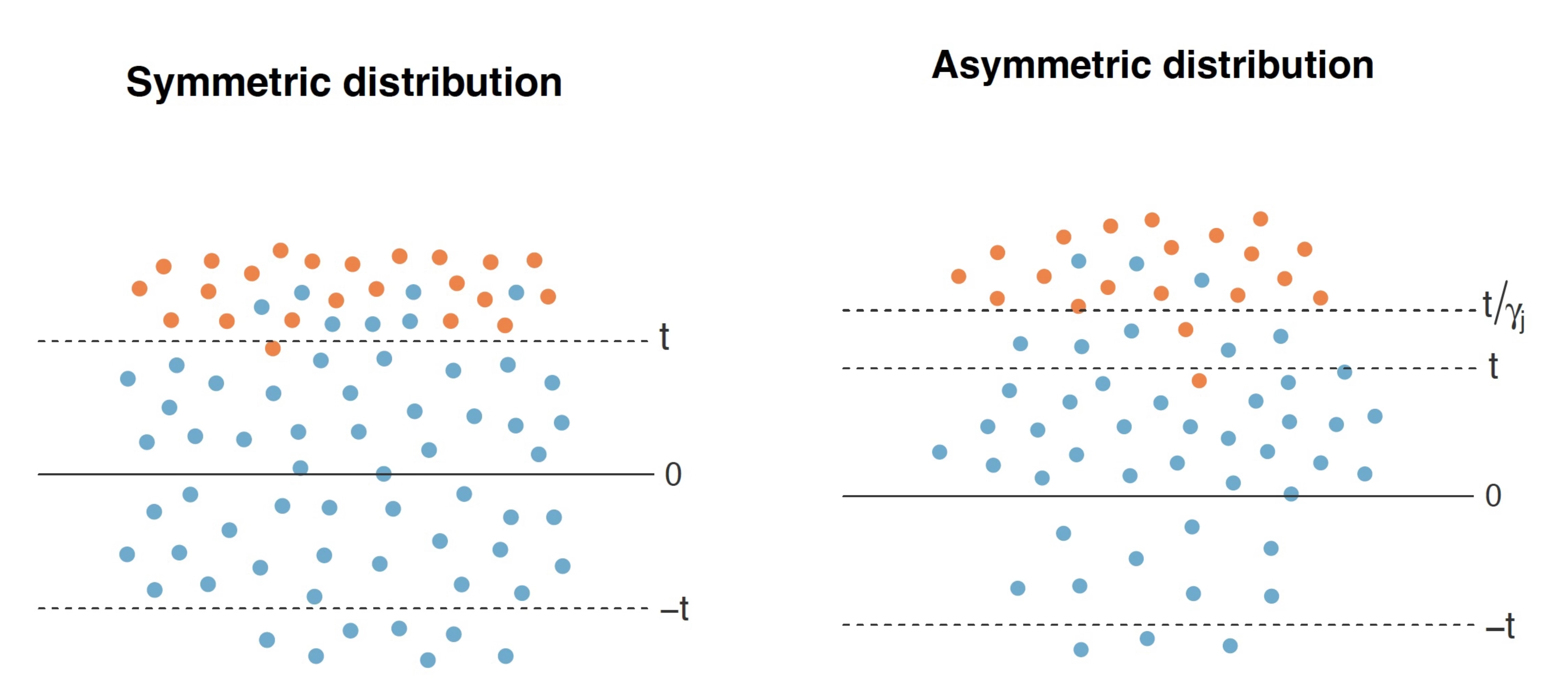}
\caption{Scatter plots of test statistics with symmetric (left) and asymmetric (right) distributions. Orange points denote {\it non-null} variables ($\mathcal{H}_1$), and blue points denote {\it null} variables ($\mathcal{H}_0$). Points above the threshold $t$ are selected as {\it non-null} ($\widehat{\mathcal{H}}_1$).} \label{scatter}
\end{figure}
Beyond the merits discussed above, the most distinctive advantage of the proposed BGM is that it avoids the strict requirement for $M_j$ to be symmetric around zero under $H_{0j}$, a key condition for both knockoff-type and mirror-type methods. To illustrate why this symmetry requirement is critical for existing approaches and how BGM overcomes it, we refer to Figure~\ref{scatter}. In the figure, each point represents a variable $X_j$: orange points indicate {\it non-null} variables ($\mathcal{H}_1$), while blue points represent {\it null} variables ($\mathcal{H}_0$). For any given threshold $t$, the points located above $t$ are selected as {\it non-null} ($\widehat{\mathcal{H}}_1$). By Equation~\eqref{FDR-select}, the FDR is estimated by the proportion of blue points above $t$ relative to the total number of points above $t$. 
In practice, however, the true label (color) of each selected variable is unknown, rendering the number of blue points above $t$ unobservable. For knockoff-type and mirror-type methods, when the test statistic $W_j$ is symmetric around zero under $H_{0j}$, it is sensible that the number of blue points above $t$ roughly equals the number below $-t$, as shown in the left panel of Figure~\ref{scatter}. This symmetry justifies using $\#\{j: W_j < -t\}$ to approximate the unobservable count $\#\{j: W_j > t \text{ and }\ j \in \mathcal{H}_0\}$ in Equation~\eqref{FDR-select}. However, when $W_j$ is asymmetric, as revealed in the right panel, this approximation breaks down, making FDR estimation unreliable for such methods. In contrast, our proposed BGM introduces pre-specified weights $\gamma_j$ to explicitly account for the asymmetry. By leveraging the exchangeability  of $W_j^{(1)}$ and $W_j^{(2)}$ of the BGM approach, 
we are able to establish that,  with an appropriately chosen $\gamma_j$, $\#\{j: M_j > t \text{ and } j \in \mathcal{H}_0\} \approx \#\{j: M_j < -t\} + \#\{j: t < M_j < t/\gamma_j\}$,
where the additional term $\#\{j: t < M_j < t/\gamma_j\}$ can be directly counted given $\gamma_j$; see Sections  \ref{BGM2.3} and \ref{section2.4} for further details. 

To clearly summarize the advantages of BGM, we compare it with four representative approaches: the model-X knockoff framework \citep{Candes2018}, the GM method \citep{Xing2023}, and the DS and multiple data splitting (MDS) methods in \cite{Dai2023}. As shown in Table~\ref{Comparison}, the BGM method relaxes assumptions on model settings, data distribution, and test statistic symmetry, while also alleviating power loss. These benefits are supported by both theoretical analysis and numerical results. Overall, BGM offers a more robust and efficient solution for FDR control in high-dimensional variable selection.

The remainder of this paper is organized as follows. Section 2 introduces the proposed BGM method for FDR control, along with a self-guiding procedure introduced to enhance its practicality. Theoretical properties, including the upper bound of FDR under null hypotheses and the power properties under alternative hypotheses, are also provided.
Section 3 presents simulation studies demonstrating that the BGM method outperforms the four competing methods listed in Table~\ref{Comparison}. Two real data analyses are provided in Section 4, revealing that our method works well empirically.
Section 5 offers concluding remarks. All technical details are relegated to the supplementary material.
 \begin{table}[!t]
	\setlength{\belowcaptionskip}{0.2cm}
	\centering
	\caption{Comparisons among BGM, \cite{Xing2023}, \cite{Dai2023} and \cite{Candes2018}. The symbol ``$\checkmark$" means applicable, and ``$\times$" stands for not applicable.}
	\label{Comparison}
	\resizebox{1.0\textwidth}{!}{
		\begin{tabular}{ccccc}
		\hline {Method}
		& \multicolumn{1}{c}{\makecell{Distribution-free data}} 
        & \multicolumn{1}{c}{\makecell{Non-significant power loss}} 
        & \multicolumn{1}{c}{\makecell{Asymmetric test statistics}} 
        & \multicolumn{1}{c}{\makecell{Generalized linear models}} \\
		\hline
		BGM &$\checkmark$ &$\checkmark$ &$\checkmark$ &$\checkmark$ \\
        \cite{Xing2023} &$\checkmark$ &$\checkmark$ &$\times$ &$\times$\\
        \cite{Dai2023} &$\checkmark$ &$\times$ &$\times$ &$\checkmark$\\
        \cite{Candes2018}  &$\times$ &$\checkmark$ &$\times$ &$\checkmark$\\
		\bottomrule
	\end{tabular}}
\end{table}
 
\section{Methodology and theoretical properties}

\subsection{Model setup and hypothesis testing}

For any matrix $\bm G$ and vector $\bm U$, let $\bm G_{i\cdot}$ and $\bm G_{\cdot j}$ denote the $i$-th row and $j$-th column of $\bm G$, respectively, and let ${\bm U}_i$ represents the $i$-th element of $\bm U$. Consider the $p$-dimensional covariate matrix with $n$ observations, denoted as $\bm X=(X_1^\top,\cdots,X_n^\top)^\top\in\mathbb{R}^{n\times p}$, and the response vector of $n$ observations, denoted as $\bm Y=(Y_1,\cdots,Y_n)^\top\in\mathbb{R}^n$. In this paper, we model the relationship between $\bm Y$ and $\bm X$ as:
\begin{equation}\label{eq:modelXY}
\mathbb{E}(\bm Y_i) = g(\bm X_{i\cdot} \beta),\quad\text{for}\quad i=1,\cdots,n, 
\end{equation}
where $g$ is a suitably chosen link function applicable to both linear and generalized linear models, and $\beta=(\beta_1,\cdots,\beta_p)^{\top}\in\mathbb{R}^p$ is the unknown coefficient vector. 
Consequently, testing hypothesis \eqref{Conditional} is equivalent to testing the following hypothesis:
\begin{equation}\label{hypo-beta}
H_{0j}: \beta_j=0 \quad \text{v.s.} \quad H_{1j}: \beta_j\not=0, \quad \text{for} \quad j =1, \cdots, p.
\end{equation}
Thus, the {\it nulls} set and the {\it non-nulls} set can be equivalently written as $\mathcal{H}_0=\{j: \beta_j=0\}$ and $\mathcal{H}_1=\{j: \beta_j\neq 0\}$, respectively. Based on \eqref{hypo-beta}, this paper aims to identify signals in $\mathcal{H}_1$, while controlling the FDR at a pre-specified level. In the remainder of this paper, without loss of generality, we assume that the covariate matrix ${\bm X}$ has been normalized.

\subsection{The feature Gaussian mirrors}

To construct the proposed BGM statistic, we first introduce two feature Gaussian mirror statistics $W_j^{(1)}$ and $W_j^{(2)}$ for $j=1,\cdots,p$.
Specifically, for each the $j$-th covariate vector $\bm X_{\cdot j}\in\mathbb{R}^{n}$, we use the Gaussian perturbation and construct two pairs of feature mirrors:
\begin{equation}\label{genXj}
    {\bm X}_{\cdot j(1)}=\left({\bm X}^{+}_{\cdot j(1)}, {\bm X}^{-}_{\cdot j(1)} \right) =\left(\bm X_{\cdot j} + \bm \zeta_j^{(1)},  \bm X_{\cdot j} - \bm \zeta_j^{(1)} \right)\in\mathbb{R}^{n\times 2},\quad\text{and}
\end{equation}
\begin{equation}\label{genXj2}
    \quad {\bm X}_{\cdot j(2)}=\left({\bm X}^{+}_{\cdot j(2)}, {\bm X}^{-}_{\cdot j(2)} \right) =\left(\bm X_{\cdot j} + \bm \zeta_j^{(2)},  \bm X_{\cdot j} - \bm \zeta_j^{(2)} \right)\in\mathbb{R}^{n\times 2},
\end{equation}
where $\bm \zeta_j^{(1)}\in\mathbb{R}^{n}$ and $\bm \zeta_j^{(2)}\in\mathbb{R}^{n}$ are independently simulated standard Gaussian random variables. It is noteworthy that Equations \eqref{genXj}–\eqref{genXj2} exclude the scalar $c_j$ required by the GM method, greatly enhancing the practicality of feature mirrors generation process. For $j=1,\cdots,p$, denote ${\bm X}_{\cdot-j}\in\mathbb{R}^{n\times(p-1)}$ as the sub-matrix of ${\bm X}$ by removing the $j$-th column, and $\beta_{-j}\in\mathbb{R}^{(p-1)\times 1}$ are the coefficient vector corresponding to ${\bm X}_{\cdot -j}$. Subsequently, we define two feature mirrors matrices: $\bm X^{j(1)}=(\bm X_{\cdot j(1)}, \bm X_{\cdot -j})\in\mathbb{R}^{n\times (p+1)}$ and  $\bm X^{j(2)}=(\bm X_{\cdot j(2)}, \bm X_{\cdot -j})\in\mathbb{R}^{n\times (p+1)}$. Correspondingly, we define two coefficient vectors: $\beta_{j}^{(1)}=(\beta^{+}_{j(1)}, {\beta}^{-}_{j(1)}, \beta_{-j})^\top\in\mathbb{R}^{p+1}$ and $\beta_{j}^{(2)}=(\beta^{+}_{j(2)}, {\beta}^{-}_{j(2)}, \beta_{-j})^\top\in\mathbb{R}^{p+1}$. 
Then, under the model \eqref{eq:modelXY}, we have
\begin{equation}\label{eq:mirror-model}
    \mathbb{E}(\bm Y_i) = g(\bm X_{i\cdot}^{j(1)} \beta_{j}^{(1)}) \quad \text{and}\quad \mathbb{E}(\bm Y_i)= g(\bm X_{i\cdot}^{j(2)} \beta_{j}^{(2)}), \quad\text{for}\quad i=1,\cdots,n.
\end{equation} 
According to Equations \eqref{genXj}--\eqref{eq:mirror-model}, we can deduce that if $j\in\mathcal{H}_0$, then  $\beta^{+}_{j(1)}=\beta^{-}_{j(1)}=\beta^{+}_{j(2)}=\beta^{-}_{j(2)}=0$. On the other hand, if $j\in\mathcal{H}_1$, we have $\beta^{+}_{j(1)}=\beta^{-}_{j(1)}=\beta^{+}_{j(2)}=\beta^{-}_{j(2)}=\beta_j/2$. {  Denote the consistent estimates of $\beta_{j}^{(1)}$ and $\beta_{j}^{(2)}$ as $\widehat\beta_{j}^{(1)}=(\widehat\beta^{+}_{j(1)}, {\widehat\beta}^{-}_{j(1)}, \widehat\beta_{-j})^\top\in\mathbb{R}^{p+1}$ and $\widehat\beta_{j}^{(2)}=(\widehat\beta^{+}_{j(2)}, {\widehat\beta}^{-}_{j(2)}, \widehat\beta_{-j})^\top\in\mathbb{R}^{p+1}$. {Then, for a pre-given nonnegative function $f(\cdot)$ satisfying $f(x)\geq 0$ for any $x\in\mathcal{R}$,} we proceed to construct two feature Gaussian mirrors  statistics, $W_j^{(1)}$ and $W_j^{(2)}$, as follows:
\begin{equation}\label{gen-ImportanceScore}
W_j^{(1)} = f\left(\widehat\beta^{+}_{j(1)},\widehat\beta^{-}_{j(1)}\right) \quad \text{and}\quad
W_j^{(2)} = f\left(\widehat\beta^{+}_{j(2)},\widehat\beta^{-}_{j(2)}\right), \quad\text{for} \quad j=1,\cdots,p.
\end{equation}

It is worth noting that the choice of the function $f(\cdot)$,  that is, the construction of the statistics $W_j^{(1)}$ and $W_j^{(2)}$, itself is an interesting research problem for all existing FDR control methods, and is beyond the scope of this paper. Further discussions are provided  below in Section \ref{section2.4}.}
As estimation methods for linear and generalized linear models are well developed, $\widehat\beta_{j}^{(1)}$ and $\widehat\beta_{j}^{(2)}$ can be obtained using a variety of existing techniques. For example, the LASSO has been extensively studied for both linear regression \citep{Tibshirani1996} and logistic regression \citep{park20071,Ravikumar2010-journal}.

\subsection{The bi-Gaussian mirrors}\label{BGM2.3}

In this section, based on the two feature Gaussian mirrors statistics $W_j^{(1)}$ and $W_j^{(2)}$ corresponding to each covariate vector $\bm X_{\cdot j}$, we propose a novel bi-Gaussian mirrors test statistic for FDR control as follows:
\begin{equation}\label{Statistics} 
M_j := M_j(\gamma_j)= W_{j}^{(1)} - \gamma_j W_{j}^{(2)}, \quad\text{for}\quad j=1,\cdots,p,
\end{equation}
where $\{\gamma_1,\cdots,\gamma_p\}$ are pre-given weights located in $(0,1]$. It is noteworthy that the weighting vector $\bm\gamma=(\gamma_1,\cdots,\gamma_p)^{\top}$ is critical for providing a reasonable upper bound on the FDR, thus playing an important role in the proposed FDR control procedure. A data-adaptive method for constructing the weighting vector $\bm\gamma$ is discussed in detail in Section \ref{section2.4}. 
According to Equations \eqref{genXj}--\eqref{genXj2}, since $\bm \zeta_j^{(1)}$,  $\bm \zeta_j^{(2)}, j=1,\ldots, p$ are independent, $W_j^{(1)}$ and $W_j^{(2)}$ are conditional independent when the design matrix ${\bm X}$ is given.  Consequently, we establish the following proposition on the conditional mutual independence of the $M_j$'s.

{
\begin{proposition}\label{proposition2} 
For any $j\neq k$, we have ${\rm Cov}(M_j, M_k\mid {\bm X}, \bm Y)=0$.
\end{proposition}

To achieve FDR control with the asymmetric statistic defined in Equation~\eqref{Statistics}, it is critical to evaluate the difference between $\mathbb{P}(M_j > s \mid \bm X, \bm Y)$ and $\mathbb{P}(M_j < -s \mid \bm X, \bm Y)$. This enables accurate estimation of the numerator in Equation~\eqref{FDR-select}, which is essential for bounding the false discovery proportion.
Under the Bi-Gaussian mirrors process, the mirror statistics $W_j^{(1)}$ and $W_j^{(2)}$ defined in \eqref{gen-ImportanceScore} 
are conditionally exchangable in the sense that $\left(W_j^{(1)}, W_j^{(2)}\right)$ and $\left(W_j^{(2)}, W_j^{(1)}\right)$ are identically distributed given ${\bm X}$ and $\bm Y$.
Leveraging this exchangeability property, the following proposition demonstrates that the difference between $\mathbb{P}(M_j > s \mid \bm X, \bm Y)$ and $\mathbb{P}(M_j < -s \mid \bm X, \bm Y)$ can be exactly evaluated.  
 
 \begin{proposition}\label{proposition1}
Assume that the distributions of $M_j$ are continuous for $j = 1, \ldots, p$. Then,
for any fixed weight vector $\bm\gamma$ and $\forall s\geq 0$, we have
 $$
\mathbb{P} \left( M_j > s \mid {\bm X}, \bm Y\right) =  \mathbb{P}\left( M_j < - s \mid {\bm X}, \bm Y \right) + \mathbb{P} \left( s\leq M_j \leq {s}/{\gamma_j}  + \left({1}/{\gamma_j} - \gamma_j \right)W_{j}^{(2)}  \mid {\bm X}, \bm Y\right).
$$
\end{proposition}
}

This proposition establishes a flexible distributional property of $M_j$, allowing its distribution to be either symmetric or asymmetric depending on the choice of weights $\gamma_j$. This flexibility effectively addresses the symmetry constraint inherent in existing knockoff-type and mirror-type statistics.
Specifically, if $\gamma_j \ne 1$, Proposition~\ref{proposition1} accounts for the discrepancy between $\mathbb{P}(M_j > s \mid \bm{X}, \bm Y)$ and $\mathbb{P}(M_j < -s \mid \bm{X}, \bm Y)$ arising from the asymmetry in the distribution of $M_j$. This enables us to 
replace the unobservable quantity $\#\{j: M_j > t \text{ and } j \in \mathcal{H}_0\}$ with the observable $\#\{j: M_j < -t\}+\#\{j: t \leq M_j \leq t/\gamma_j + \left(1/\gamma_j - \gamma_j \right) W_j^{(2)}\}$ when estimating FDR using Equation~\eqref{FDR-select}, for any given threshold $t$. 
In contrast, if $\gamma_j = 1$,  Proposition~\ref{proposition1} shows
$\mathbb{P}(M_j > s \mid \bm{X}, \bm Y) = \mathbb{P}(M_j < -s \mid \bm{X}, \bm Y)$,
indicating that $M_j$ is symmetric around zero. This symmetry allows us to approximate $\#\{j: M_j > t \text{ and } j \in \mathcal{H}_0\}$ by $\#\{j: M_j < -t\}$.
Additionally, we have that if $\gamma_j\rightarrow 0$, then $\lim\limits_{\gamma_j \rightarrow 0}\mathbb{P} \left( s\leq M_j \leq {s}/{\gamma_j}  + \left({1}/{\gamma_j} - \gamma_j \right)W_{j}^{(2)}  \mid {\bm X}, \bm Y\right)=\mathbb{P} \left( M_j > s \mid {\bm X}, \bm Y\right)$, which implies that $\mathbb{P}\left( M_j < - s \mid {\bm X}, \bm Y \right)\to 0$ and hence $M_j$ tends to be large and positive in this regime.
These discussions motivate the idea that if we can construct a well-calibrated $\gamma_j$ such that $\gamma_j \to 1$ for $j \in \mathcal{H}_0$, but $\gamma_j \to 0$ or becomes significantly smaller than 1 for $j \in \mathcal{H}_1$, then we can expect to achieve effective separation between {\it non-null} and {\it null} covariates using $M_j$ values. This aspect is thoroughly investigated in the following section.


\subsection{A self-guiding procedure for FDR control}\label{section2.4}

As discussed in Section 2.3, a well-constructed $\gamma_j$ is expected to satisfy $\gamma_j \to 1$ for $j \in \mathcal{H}_0$, while $\gamma_j \to 0$ or becomes significantly smaller than 1 for $j \in \mathcal{H}_1$. To achieve this behavior, we first introduce the following assumption to properly define a functional form of $\gamma_j$.

\begin{assumption}[Weight functionalization]\label{assumption2}
Suppose that $\gamma_j = h(\vert\beta_j\vert)$ for $j=1,\cdots,p$, where $h(\cdot)$ is a monotone function satisfying $\lim_{m\rightarrow 0^{+}}h(m)=1$ and $\lim_{m\rightarrow \infty}h(m)=0$.
\end{assumption}


Based on Assumption \ref{assumption2}, in this paper, we consider the following weighting function:
\begin{equation}\label{gammafun}\nonumber
 h(m) = \frac{1}{(m+1)^{\kappa}},\quad\text{for}\quad m\geq0,
\end{equation}
where $\kappa\geq 1$ is a fixed constant.
Then, for any consistently estimated coefficient $\widehat{\beta}_j$, the corresponding estimated weight is:
\begin{equation}\label{hatgamma}
    \widehat{\gamma}_j(\kappa) :=h(\vert\widehat{\beta}_j\vert)= \frac{1}{(\vert\widehat{\beta}_j\vert + 1)^{\kappa}}, \quad\text{for}\quad j=1,\cdots,p.
\end{equation}
{ Finally, we define the self-guiding bi-Gaussian mirrors test statistic as:
\begin{equation}\label{hatMj}
    \widehat{M}_j(\kappa) = W_j^{(1)} - \widehat{\gamma}_j(\kappa) W_j^{(2)}, \quad \text{for}\quad j=1,\cdots,p.
\end{equation}
To facilitate the theoretical analysis of $\widehat{M}_j(\kappa)$, in this work, we construct the feature Gaussian mirror statistics $W_j^{(1)}$ and $W_j^{(2)}$ in Equation~\eqref{hatMj} as follows:
 \begin{equation}\label{ImportanceScore}
W_j^{(1)} = 4\left\vert\widehat\beta^{+}_{j(1)}\widehat\beta^{-}_{j(1)}\right\vert \quad \text{and}\quad
W_j^{(2)} = 4\left\vert\widehat\beta^{+}_{j(2)}\widehat\beta^{-}_{j(2)}\right\vert, \quad\text{for} \quad j=1,\cdots,p.
\end{equation}

We remark that the choice of the nonnegative function $f(\cdot)$ in \eqref{gen-ImportanceScore} is not unique. Any nonnegative continuous function $f$ is valid provided it serves as a score that effectively captures the importance of the variables. 
For example, setting $W_j^{(1)}=\vert\widehat{\beta}^{+}_{j(1)}\vert + \vert\widehat{\beta}^{-}_{j(1)}\vert$ and $W_j^{(2)}=\vert\widehat{\beta}^{+}_{j(2)}\vert + \vert\widehat{\beta}^{-}_{j(2)}\vert$ also meets the requirement.   
}

According to Assumption \ref{assumption2}, a consistently estimated $\widehat\beta_j$ is critical for obtaining a reasonable estimate of $\gamma_j$, and this is crucial for the performance of $\widehat{M}_j(\kappa)$ in controlling the FDR. Therefore, to ensure a well-estimated $\widehat{\gamma}_j(\kappa)$ and to establish the theoretical properties of $\widehat{M}_j(\kappa)$, we impose the following assumption on $\widehat{\beta}$.

\begin{assumption}[Estimates consistency]\label{assumption3}
Suppose that there exists a non-negative sequence $\delta_{n,p}\rightarrow0$, associated with $n$ and $p$, such that the estimated linear coefficients $\widehat{\beta}_{j}^{(1)}$ and  $\widehat{\beta}_{j}^{(2)}$ corresponding to the feature Gaussian mirrors statistics $W_j^{(1)}$ and $W_j^{(2)}$, respectively, satisfy 
$
\sup_{j=1,\cdots,p}\|\widehat{\beta}_{j}^{(m)}-\beta_{j}^{(m)}\|_{\infty}=O_{P}(\delta_{n,p})
$
for  $m=1,2$.
In addition, the initial estimated linear coefficient vector $\widehat{\beta}$ also satisfies
$\|\widehat{\beta}-\beta\|_{\infty}=O_{P}(\delta_{n,p})$.
\end{assumption}

This assumption ensures the consistency of the estimated linear coefficients $\widehat{\beta}_{j}^{(1)}$, $\widehat{\beta}_{j}^{(2)}$, and $\widehat{\beta}$. Based on this, we present the following lemma, which establishes the consistency of the estimated weights $\widehat{\gamma}_j(\kappa)$ under the null hypothesis $H_{0j}$ for $j =1, \cdots, p$.
{ 
\begin{lemma}\label{lemma1}
Under Assumptions \ref{assumption2}--\ref{assumption3}, $\sup_{j\in\mathcal{H}_0}\left(1 - \widehat{\gamma}_j(\kappa)\right)=o_{P}(1)$ holds for any given $\kappa \ge 1$.
\end{lemma}}

This lemma indicates that $\widehat{\gamma}_j(\kappa)\to 1$ for any $j\in\mathcal{H}_0$, which follows intuitively from the form of Equation \eqref{hatgamma}, as $\widehat{\beta}_j\to 0$ under Assumption \ref{assumption3}. We now present the following lemma concerning the boundedness property of the self-guiding BGM test statistics \eqref{hatMj}.

{ 

\begin{lemma}\label{lemma2}
Under Assumptions \ref{assumption2}--\ref{assumption3}, we have,  uniformly for all  $j\in \mathcal{H}_0$, $\widehat{M}_j(\kappa)$ is bounded with probability tending to 1.

\end{lemma}
}
This boundedness property ensures that, under suitable signal strength assumptions, a cutoff value can be identified to distinguish the {\it null} and {\it non-null} covariates. In the following, we describe the procedure for determining this cutoff value.

To achieve FDR control at a given level $q \in (0, 1)$, a commonly used data-adaptive procedure determines an appropriate threshold $t$ by ranking all $\widehat{M}_j(\kappa)$ values in ascending order, identifying a cutoff value $t = \tau_q(\kappa)$, and selecting the set of {\it non-nulls} as $\widehat{\mathcal{H}}_1 = \{j : \widehat{M}_j(\kappa) > \tau_q(\kappa)\}$ \citep{Barber2015, Xing2023, Dai2022}. According to Proposition~\ref{proposition1}, for any $s \ge 0$, we have
\begin{align}
\mathbb{P} \left(  \widehat{M}_j(\kappa)> s \mid {\bm X}, \bm Y\right) = &\mathbb{P}\left( \widehat{M}_j(\kappa)<-s  \mid {\bm X}, \bm Y\right) + \mathbb{P} \left( s\leq  \widehat{M}_j(\kappa) < \frac{s}{\widehat{\gamma}_j(\kappa)}\mid {\bm X}, \bm Y\right)\nonumber \\
& +  \mathbb{P} \left(\frac{s}{\widehat{\gamma}_j(\kappa)}\leq \widehat{M}_j(\kappa) \leq \frac{s}{\widehat{\gamma}_j(\kappa)} + \left(\frac{1}{\widehat{\gamma}_j(\kappa)} - \widehat{\gamma}_j(\kappa) \right)W_{j}^{(2)}  \mid {\bm X}, \bm Y\right)\nonumber\\
 := & \widehat{R}_1(s) + \widehat{R}_2(s) + \widehat{R}_3(s),\label{Theo-Mj}
\end{align}
which implies that, for any given threshold $t$, we have
\begin{align}
\#\{j\in\mathcal{H}_0: \widehat{M}_j(\kappa) > t\} \approx & \#\{j\in\mathcal{H}_0: \widehat{M}_j(\kappa) < -t \} + \#\left\{j\in\mathcal{H}_0:  t\leq \widehat{M}_j(\kappa) < \frac{t}{\widehat{\gamma}_j(\kappa)} \right\}  \nonumber \\
& +  \#\left\{j\in\mathcal{H}_0:  \frac{t}{\widehat{\gamma}_j(\kappa)}\leq \widehat{M}_j(\kappa) \leq \frac{t}{\widehat{\gamma}_j(\kappa)}  + \left(\frac{1}{\widehat{\gamma}_j(\kappa)} - \widehat{\gamma}_j(\kappa) \right)W_{j}^{(2)} \right\}\nonumber\\
\leq & \#\{j: \widehat{M}_j(\kappa) < -t \} + \#\left\{j:  t\leq \widehat{M}_j(\kappa) < \frac{t}{\widehat{\gamma}_j(\kappa)} \right\} \nonumber\\
& + \#\left\{j:  \frac{t}{\widehat{\gamma}_j(\kappa)}\leq \widehat{M}_j(\kappa) \leq \frac{t}{\widehat{\gamma}_j(\kappa)}  + \left(\frac{1}{\widehat{\gamma}_j(\kappa)} - \widehat{\gamma}_j(\kappa) \right)W_{j}^{(2)} \right\}\nonumber\\
:=& \widehat{V}_1(t) + \widehat{V}_2(t) + \widehat{V}_3(t).\label{Emp-Mj}
\end{align}

In the following, we present a lemma that simplifies the expression in Equation~\eqref{Theo-Mj}. Building on this result, we then define a well-performing, data-driven cutoff value tailored to the self-guiding BGM test statistic.

\begin{lemma}\label{lemma3} Under Assumptions \ref{assumption2}--\ref{assumption3} and $H_{0j}: \beta_j =0 $, we have $\widehat{R}_3(s) = o_{P}\big\{\widehat{R}_2(s)\big\}$.
\end{lemma}
This lemma indicates that $\widehat{V}_3(t)=o_P\big\{\widehat{V}_2(t)\big\}$ with respect to its empirical form in Equation~\eqref{Emp-Mj}.
Consequently, for any given FDR level $q\in(0,1)$, we define the cutoff value corresponding to $\widehat{M}_j(\kappa)$ as follows:
\begin{align}
\tau_q(\kappa) &=\min_t\left\{t>0: \frac{\widehat{V}_1(t) + \widehat{V}_2(t)}{\#\{j: \widehat{M}_j(\kappa) > t\}\vee 1} \leq q \right\}\nonumber\\
&=\min_t\left\{t>0:  \frac{\#\{j: \widehat{M}_j(\kappa) < - t \}  +  \#\left\{j:  t\leq \widehat{M}_j(\kappa) < {t}/{\widehat{\gamma}_j(\kappa)}   \right\} }{\#\{j: 
\widehat{M}_j(\kappa) > t \}\vee 1} \leq q \right\}.\label{tauq}
\end{align}
The {\it non-nulls} set is finally selected as: $\widehat{\mathcal{H}}_1=\{j: \widehat{M}_j(\kappa)> \tau_q(\kappa)\}$. 
For clarity and ease of application, we present our self-guiding FDR control procedure in Algorithm \ref{Algorithm1}.
In the following, we present the theoretical results regarding the FDR control and the asymptotic power of the proposed self-guiding BGM procedure, respectively. To establish the FDR control property, we first introduce the following assumption.

\begin{assumption}[Weak dependence]\label{Weakdependence} Let $p_0$ represent the number of covariates in $\mathcal{H}_0$. Note that $\widehat{M}_j(\kappa)$'s are continuous random variables for any fixed $\kappa\geq 1$. Suppose that, for any given threshold $t\in\mathbb{R}$, there exist constants $c> 0$ and $\alpha \in(0,2)$ such that
$$\text{Var}\left(\sum_{j\in\mathcal{H}_0}\mathbb{I}\big\{\widehat{M}_j(\kappa)>t\big\}\right)\leq cp_0^\alpha.$$
\end{assumption}
{By Proposition \ref{proposition2}, we have that, conditioned on $\bm X$ and $\bm Y$, $\widehat{M}_j(\kappa)$ and $\widehat{M}_k(\kappa)$ are independent for any $j\neq k$. Therefore, if $\bm X_j$ is independent or weakly correlated to some extent for $j \in \mathcal{H}_0$, Assumption \ref{Weakdependence} is satisfied.} It is worth noting that this assumption only restricts the correlations among the {\it null} features, regardless of the correlations associated with the relevant features. A similar assumption was considered in \cite{Dai2022}.

\IncMargin{1em}
\begin{algorithm}\footnotesize
\SetKwInOut{Input}{Input}
\SetKwInOut{Output}{Output}
\caption{The self-guiding BGM procedure for FDR control}
\label{Algorithm1}
	
\Input{The design matrix ${\bm X}\in\mathbb{R}^{n\times p}$, the response vector $\bm Y\in\mathbb{R}^n$, the weighting parameters $1\leq \kappa_1 <\kappa_2<\cdots<\kappa_d$, and the FDR level $q\in(0,1)$. } 

Solve the estimation problem under model \eqref{eq:modelXY} and get the initial regression coefficient estimate $\widehat{\beta}=(\widehat{\beta}_1,\cdots,\widehat{\beta}_p)^{\top}$.

For each covariates vector $\bm X_{\cdot j}$, generate its feature Gaussian mirrors $\bm X_{\cdot j(1)}$ and $\bm X_{\cdot j(2)}$ using Equation \eqref{genXj}:
$$\bm X_{\cdot j(m)}=
\left({\bm X}^{+}_{\cdot j(m)}, {\bm X}^{-}_{\cdot j(m)} \right) =\left(\bm X_{\cdot j} + \bm \zeta_j^{(m)},  \bm X_{\cdot j} - \bm \zeta_j^{(m)} \right),\quad\text{for}\quad m=1,2,
$$
where $\bm \zeta_j^{(1)}$ and $\bm \zeta_j^{(2)}$ are independently simulated standard Gaussian random variables.

Solve the estimation problem under model \eqref{eq:mirror-model} and calculate the feature Gaussian mirrors statistics $W_j^{(1)}$ and $W_j^{(2)}$ using Equation \eqref{ImportanceScore}: 
$$
W_j^{(1)} = 4\left\vert\widehat{\beta}^{+}_{j(1)}\widehat{\beta}^{-}_{j(1)}\right\vert \quad \text{and}\quad
W_j^{(2)} = 4\left\vert\widehat{\beta}^{+}_{j(2)}\widehat{\beta}^{-}_{j(2)}\right\vert, \quad \text{for}\quad j=1,\cdots,p.
$$

\For{$\kappa$ {\rm in} $\{\kappa_1,\cdots,\kappa_d\}$ }
{
Determine the weights using Equation \eqref{hatgamma}:
$$
\widehat{\gamma}_j(\kappa) = \frac{1}{(\vert\widehat{\beta}_j\vert + 1)^{\kappa}},\quad \text{for}\quad j=1,\cdots,p.
$$

Calculate the self-guiding BGM test statistics using Equation \eqref{hatMj}:
$$ \widehat{M}_j(\kappa) = W_j^{(1)} - \widehat{\gamma}_j(\kappa) W_j^{(2)},\quad \text{for}\quad j=1,\cdots,p.$$

Calculate the cutoff value $\tau_q(\kappa)$ at the given FDR level using Equation \eqref{tauq}:
 $$
\tau_q(\kappa)=\min_t\left\{t>0:  \frac{\#\{j: \widehat{M}_j(\kappa) < - t \}  +  \#\left\{j:  t\leq \widehat{M}_j(\kappa) < {t}/{\widehat{\gamma}_j(\kappa)}   \right\} }{\#\{j: \widehat{M}_j(\kappa) > t \}\vee 1} \leq q \right\}.
$$

Get the selected set $\widehat{\mathcal{H}}_1(\kappa)=\{j: \widehat{M}_j(\kappa)> \tau_q(\kappa)\}$.
}

Compute $\kappa_{\max} = \arg\max\limits_{\kappa\in\{\kappa_1,\cdots,\kappa_d\}}\vert\widehat{\mathcal{H}}_1(\kappa)\vert.$

\Output{The final selected set $\widehat{\mathcal{H}}_1=\widehat{\mathcal{H}}_1( \kappa_{\max} )$. }
\end{algorithm} 
 \DecMargin{1em} 


\begin{theorem}\label{theorem-null}
 Suppose Assumptions \ref{assumption2}–\ref{Weakdependence} hold. Let $\widehat{M}_j(\kappa)$ represent the self-guiding BGM statistics for $j=1, \cdots, p$, where $\kappa \geq 1$ is a fixed constant as defined in \eqref{hatgamma}. The selected non-null variables are given by $\widehat{\mathcal{H}}_1 =\{j: \widehat{M}_j(\kappa)> \tau_q(\kappa)\}$, where $\tau_q(\kappa) > 0$ is the cutoff value at any given level $q \in (0, 1)$ as defined in \eqref{tauq}.  Let $p_0$ denote the number of covariates in $\mathcal{H}_0$. 
 Then, for any given $\kappa$, as $p_0\to\infty$ with $p\to\infty$, we have
\begin{equation}\label{Null-FDRkappa}\nonumber
\mathrm{FDR}(\kappa)=\mathbb{E}\left[\frac{\#\{j: j \in \widehat{\mathcal{H}}_1(\kappa) \text{ and } j \in \mathcal{H}_0 \}}{\#\{j: j \in \widehat{\mathcal{H}}_1(\kappa) \} \vee 1}\right]\le q. \end{equation}

\end{theorem}

This theorem implies that we can obtain the estimated {\it non-nulls} set $\widehat{\mathcal{H}}_1$ while effectively controlling the FDR level using our proposed method. In the following, to establish the asymptotic power of the proposed self-guiding BGM statistic, we introduce an additional assumption given below.

\begin{assumption} [Signal strength]\label{assumption-signal}
Suppose that $\min_{j\in\mathcal{H}_1}|\beta_j|/\delta_{n,p}\to\infty$,
where $\delta_{n,p}\to 0$ is a non-negative sequence associated with $n$ and $p$, as defined in Assumption \ref{assumption3}.
\end{assumption}

This assumption requires the signal strength to exceed the coefficient estimation error $\delta_{n,p}$.  Together with Assumption \ref{assumption3}, this implies that the coefficient estimates $\widehat\beta_j$, $\widehat\beta_j^{(1)}$, and $\widehat\beta_j^{(2)}$ for $j \in \mathcal{H}_1$ are expected to be larger than those for $j \in \mathcal{H}_0$. Thus, Assumption \ref{assumption-signal} plays a key role in powerfully distinguishing between the {\it null} and {\it non-null} covariates.

\begin{theorem}\label{theorem-alter}
 Suppose Assumptions \ref{assumption2}, \ref{assumption3} and \ref{assumption-signal} hold. Let $p_1$ denote the number of covariates in $\mathcal{H}_1$. Under the same other statements outlined in Theorem~\ref{theorem-null}, for any given $\kappa$ and as $p_1\to\infty$ with $p\to\infty$, we have 
\begin{equation*}
\mathrm{Power}(\kappa)=\mathbb{E}\left[\frac{\#\{j: j \in \widehat{\mathcal{H}}_1(\kappa) \text{ and } j \in \mathcal{H}_1 \}}{\#\{j: j \in \mathcal{H}_1(\kappa) \} \vee 1}\right]\to 1. \end{equation*} 
\end{theorem}

This theorem demonstrates that the proposed self-guiding BGM test statistic can powerfully against the alternatives, a property that is not theoretically established for many existing selection methods \citep{Barber2015,Xing2023}. Therefore, Theorem~\ref{theorem-alter} enhances the theoretical value of our proposed method.

\section{Simulation studies}

In this section, we conduct numerical experiments to demonstrate the finite-sample performance of the proposed self-guiding BGM method. 
In addition, to evaluate the superiority of our approach, we compare it with four established FDR control methods, all tailored for variable selection in high-dimensional data: (i) the Knockoff procedure from \cite{Candes2018}; (ii) the GM procedure from \cite{Xing2023}; (iii) the DS and MDS methods from \cite{Dai2023}.
To demonstrate the flexibility of our approach, we conduct simulations under both linear regression and logistic regression models. It is important to mention that since the GM procedure was developed for linear models, it is excluded from our comparisons under the logistic regression setting.
In the following, we provide a detailed description of the data generation, model settings, implementation of each method, and a summary of the simulation results.

{\bf Covariate generation:} We follow a similar setup as in \cite{Dai2023} to generate the dependent covariate matrix $\bm X\in\mathbb{R}^{n\times p}$. Let $p^{\prime}=p/10$.
Define $\bm\Sigma_{B}$ as a blockwise diagonal matrix consisting of 10 identical unit-diagonal Toeplitz matrices. Each block along the diagonal is structured as:
$$
\left[\begin{array}{cccccc}
1 & \frac{(p^{\prime}-2)\rho}{p^{\prime}-1} & \frac{(p^{\prime}-3)\rho}{p^{\prime}-1} & \cdots & \frac{\rho}{p^{\prime}-1} & 0 \\
\frac{(p^{\prime}-2)\rho}{p^{\prime}-1} & 1 & \frac{(p^{\prime}-2)\rho}{p^{\prime}-1}  & \cdots & \frac{2\rho}{p^{\prime}-1}  & \frac{\rho}{p^{\prime}-1}  \\
\vdots & & \ddots & & & \vdots \\
0 & \frac{\rho}{p^{\prime}-1}   &  \frac{2\rho}{p^{\prime}-1}   & \cdots & \frac{(p^{\prime}-2)\rho}{p^{\prime}-1} & 1
\end{array}\right]\in\mathbb{R}^{p^\prime\times p^\prime}.
$$ 
We refer to  $\rho\in (0,1)$  as the correlation factor within each block of $\bm\Sigma_{B}$. Then, each row of the design matrix $\bm X$ is independently generated from the multivariate normal distribution $N(0,\bm\Sigma_B)$, where the correlation factor $\rho$ takes values from the set $\{0.2, 0.3, 0.4, 0.5, 0.6\}$, allowing us to simulate covariates with varying levels of correlation.

{\bf Model settings:} We consider both the linear regression model and the logistic regression model as follows:

\begin{itemize}

\item[(1)] Simulate the response vector $\bm Y\in\mathbb{R}^n$ using the following linear regression model:
$$\bm Y = \bm{X}\beta+ \varepsilon,$$  
where $\varepsilon \sim N(0, \bm{I}_n)$ is the random error vector, with $\bm{I}_n$ denoting the $n$-dimensional identity matrix.
The sample size is set to $n = 400$, the dimension $p = 1000$, and the number of {\it non-null} covariates is $\vert\mathcal{H}_1\vert = 50$. The {\it non-null} covariates are randomly selected from $\{1,\cdots,p\}$. Then,
we introduce $\delta>0$ as the signal amplitude and independently generate the {\it non-null} components $\beta_j$ for $j\in\mathcal{H}_1$ from the uniform distribution ${\rm Unif}(\delta\sqrt{\log(p)/n}, \delta\sqrt{\log(p)/n}+0.2)$. The values of $\delta$ are varied from $2.5$ to $3.5$ in increments of $0.25$. For the {\it null} covariates  $j\in\mathcal{H}_0$, we set $\beta_j=0$.
 
\item[(2)] Generate the response vector $\bm Y\in\mathbb{R}^n$ using the following logistic regression model:
$$
\bm Y_i\mid\bm X_{i\cdot} \sim {\rm Bernoulli}\left(\frac{ e^{\bm X_{i\cdot}\beta} }{1 + e^{\bm X_{i\cdot}\beta}} \right).
$$
In this setup, the sample size is set to $n=600$, the dimension $p=1000$, and the number of {\it non-null} covariates is $\vert\mathcal{H}_1\vert = 20$. The {\it non-null} covariates are randomly selected from $\{1,\cdots,p\}$. For each {\it non-null} covariate $j\in\mathcal{H}_1$, the corresponding coefficient is set as $\beta_j = \delta\sqrt{\log(p)/n}$, where $\delta$ varies from $9.0$ to $11.0$ in increments of $0.5$. The remaining components of $\beta$ are set to zero, i.e. $\beta_j=0$ for $j\in\mathcal{H}_0$.
\end{itemize}

\begin{figure}[H]
\centering
\includegraphics[scale=0.25]{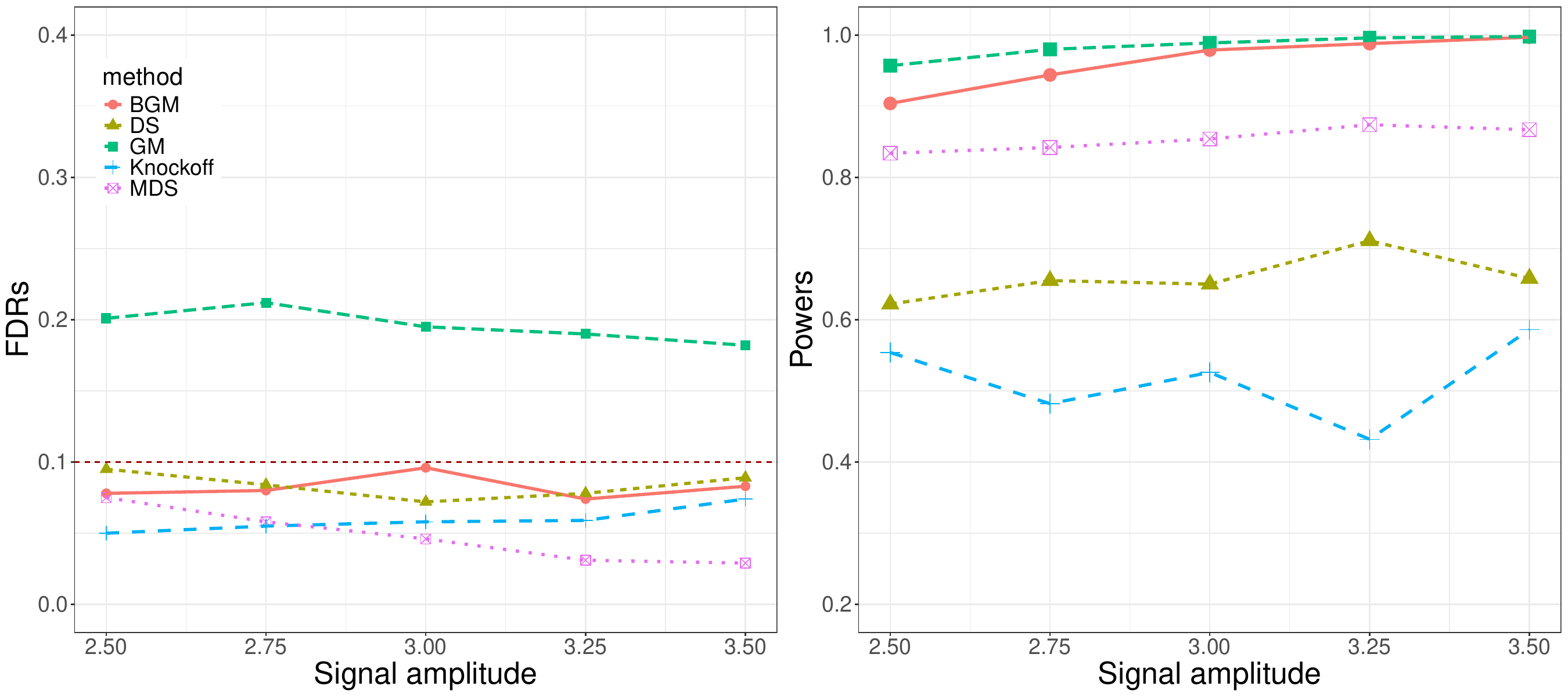}
\caption{FDRs and powers under the linear regression model, with the signal amplitude $\delta$ varying from $2.5$ to $3.5$ in increments of $0.25$ and the correlation factor $\rho$ fixed at 0.5.}
\label{Figure1}
\end{figure}

\begin{figure}[H]
\centering
\includegraphics[scale=0.25]{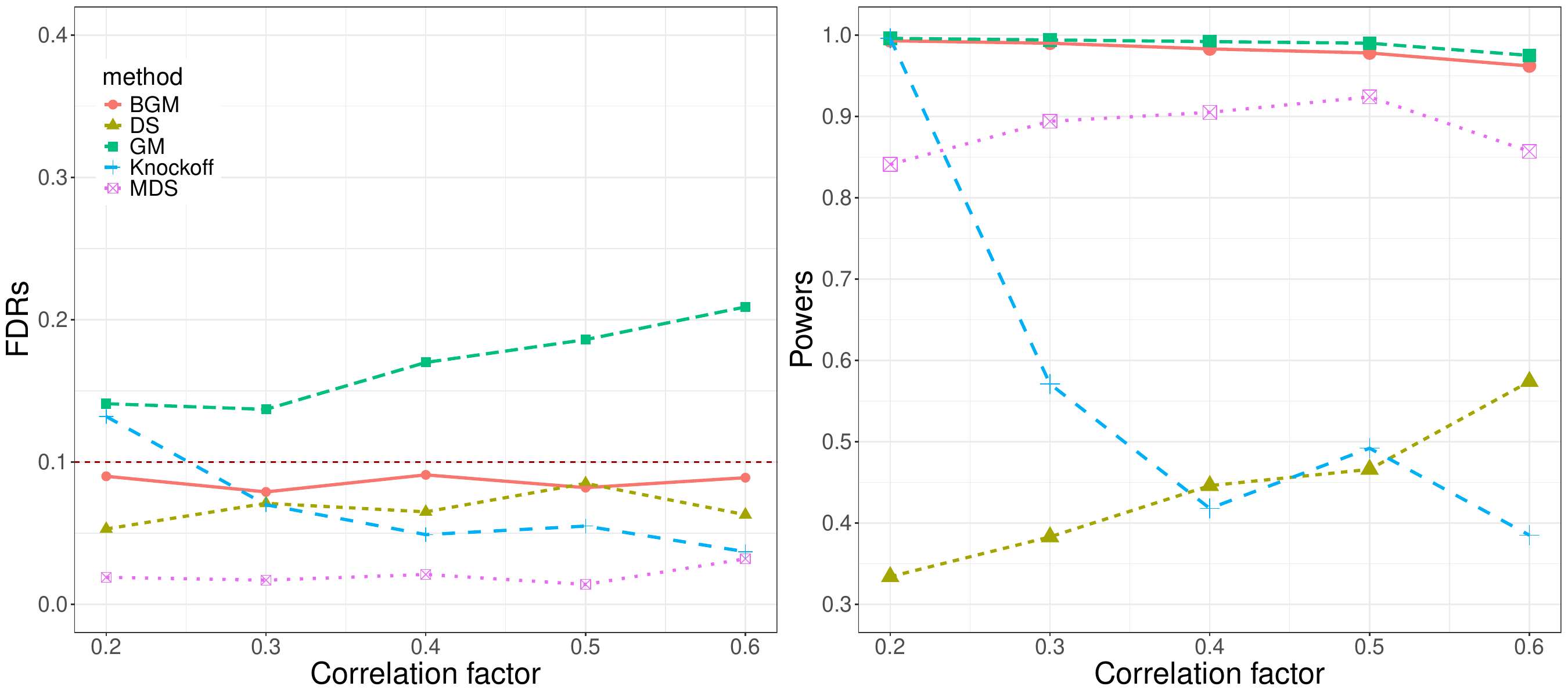}
\caption{FDRs and powers under the linear regression model, with the varying correlation factors $\rho\in \{0.2, 0.3, 0.4, 0.5, 0.6\}$ and the signal amplitude $\delta$ fixed at 3.0.}
\label{Figure2}
\end{figure}

\begin{figure}[H]
\centering
\includegraphics[scale=0.25]{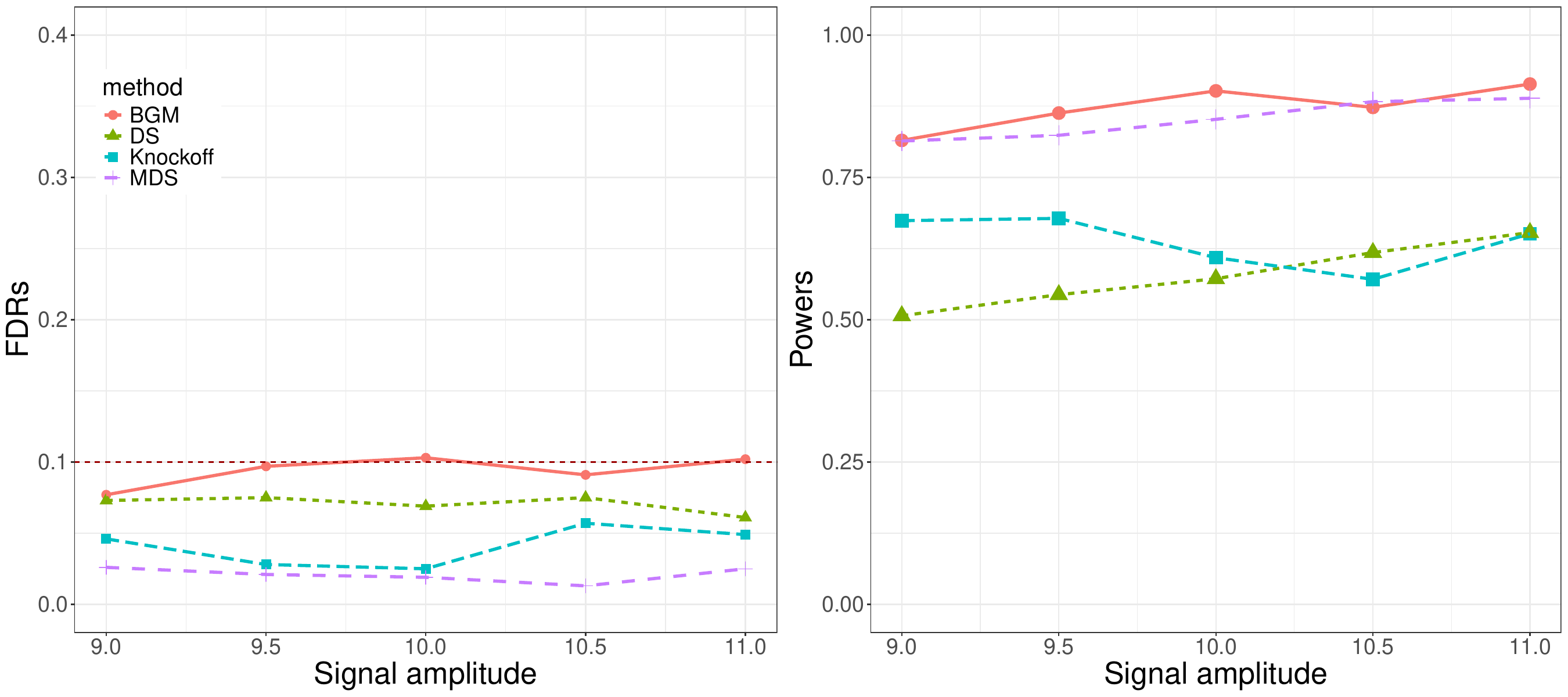}
\caption{FDRs and powers under the logistic regression model, with the signal amplitude $\delta$ varying from $9.0$ to $11.0$ in increments of $0.5$ and the correlation factor $\rho$ fixed at 0.2.}
\label{Figure3}
\end{figure}

{\bf  Implementation of each method:} We implement our self-guiding BGM procedure following Algorithm \ref{Algorithm1}. The weighting parameter $\kappa$ is set to take values in $\{1, 2, \cdots, 10\}$. For both the linear and logistic regression models, we use the LASSO estimator to obtain $\widehat{\beta}_j$.
The four competing methods are implemented using the R packages ``knockoff'' and ``GM'', as well as the R code provided by \cite{Dai2023} on GitHub for the DS and MDS methods, respectively. The designated FDR control level is set at 
$q=0.1$. The simulation results are based on 50 realizations for the linear model and 100 realizations for the logistic regression model, and they are summarized in Figures \ref{Figure1}-\ref{Figure3}, respectively.

{\bf Results summary:} According to Figures \ref{Figure1}-\ref{Figure3}, we draw the following two conclusions: (i) For FDR control under the linear regression, The BGM method outperforms the GM method, as well as the Knockoff method when $\rho=0.2$. Additionally, the BGM method is generally comparable to the DS method and surpasses it when $\delta=3.0$ and $\rho=0.4$, while it is less effective than the MDS method. For the logistic regression,
although our method is somewhat less effective at controlling FDR compared to the DS, MDS, and Knockoff methods, it still successfully maintains the FDR within the target level. (ii) The BGM method demonstrates greater power against alternatives than all competing methods, except for the GM method, which performs comparably under the linear regression.

In sum, the proposed BGM procedure surpasses the four competing methods in terms of either FDR control or empirical power across both linear and logistic regression models. As a result, it achieves a favorable balance between FDR control and test power, demonstrating greater practicality compared to the alternative methods.

\section{Real data analyses}

To demonstrate the practical utility of our self-guiding BGM procedure, we apply it to two real datasets and compare the results with those obtained from the four competing methods considered in the simulation studies.
The first dataset is related to Human Immunodeficiency Virus Type 1 (HIV-1). The analysis for this dataset is performed under the linear regression model, with the objective of identifying mutations in HIV-1 associated with drug resistance, an objective also investigated by \cite{Dai2022}. The second dataset involves single-cell RNA sequencing (scRNA-seq) data from human breast cancer cells, as studied in \cite{hoffman2020single}. The goal is to identify relevant genes associated with the glucocorticoid response in this cell line, which was also explored in \cite{Dai2023}. We analyze this dataset using a logistic regression model.
The GM approach is excluded from the comparison in the scRNA-seq data analysis, as it is restricted to linear models.

\subsection{Human Immunodeficiency Virus Type 1}

This dataset contains drug resistance measurements and genotype information from HIV-1 samples and is available at \url{https://hivdb.stanford.edu/pages/published_analysis/genophenoPNAS2006/}. According to \cite{rhee2006genotypic}, it features separate datasets for resistance measurements of seven protease inhibitors (PIs) drugs, six targeting nucleoside reverse transcriptase inhibitors (NRTIs) drugs, and three targeting nonnucleoside reverse transcriptase inhibitors (NNRTIs) drugs. The dataset sizes are presented as follows.

\begin{figure}[H]
\centering
\includegraphics[scale=0.5]{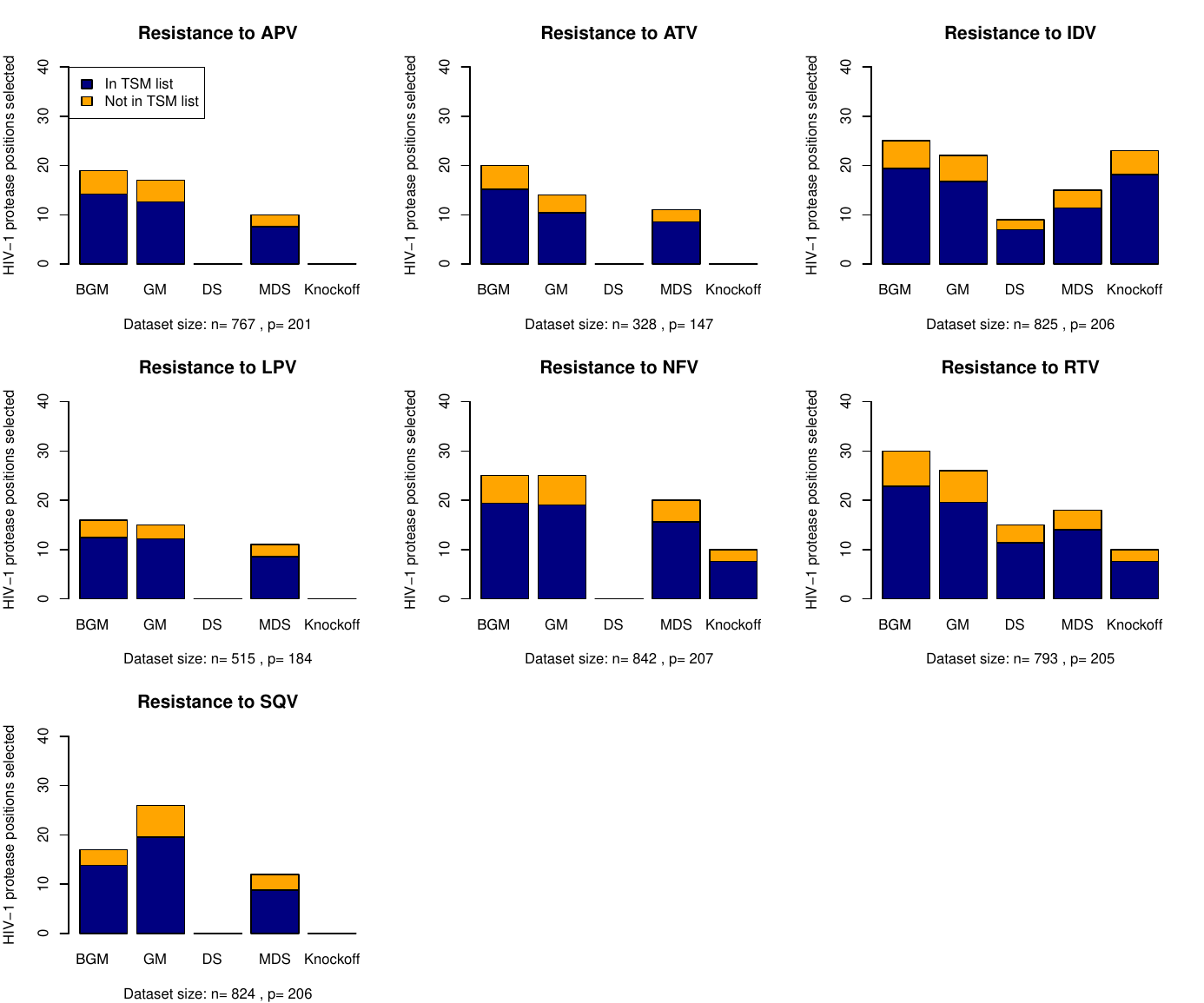}
\caption{Numbers of the discovered mutations for the seven PI drugs. The dark blue and orange bars represent the numbers of true and false positives, respectively. The designated FDR control level is $q = 0.1$.}
\label{PI}
\end{figure}

\begin{figure}[H]
\centering
\includegraphics[scale=0.5]{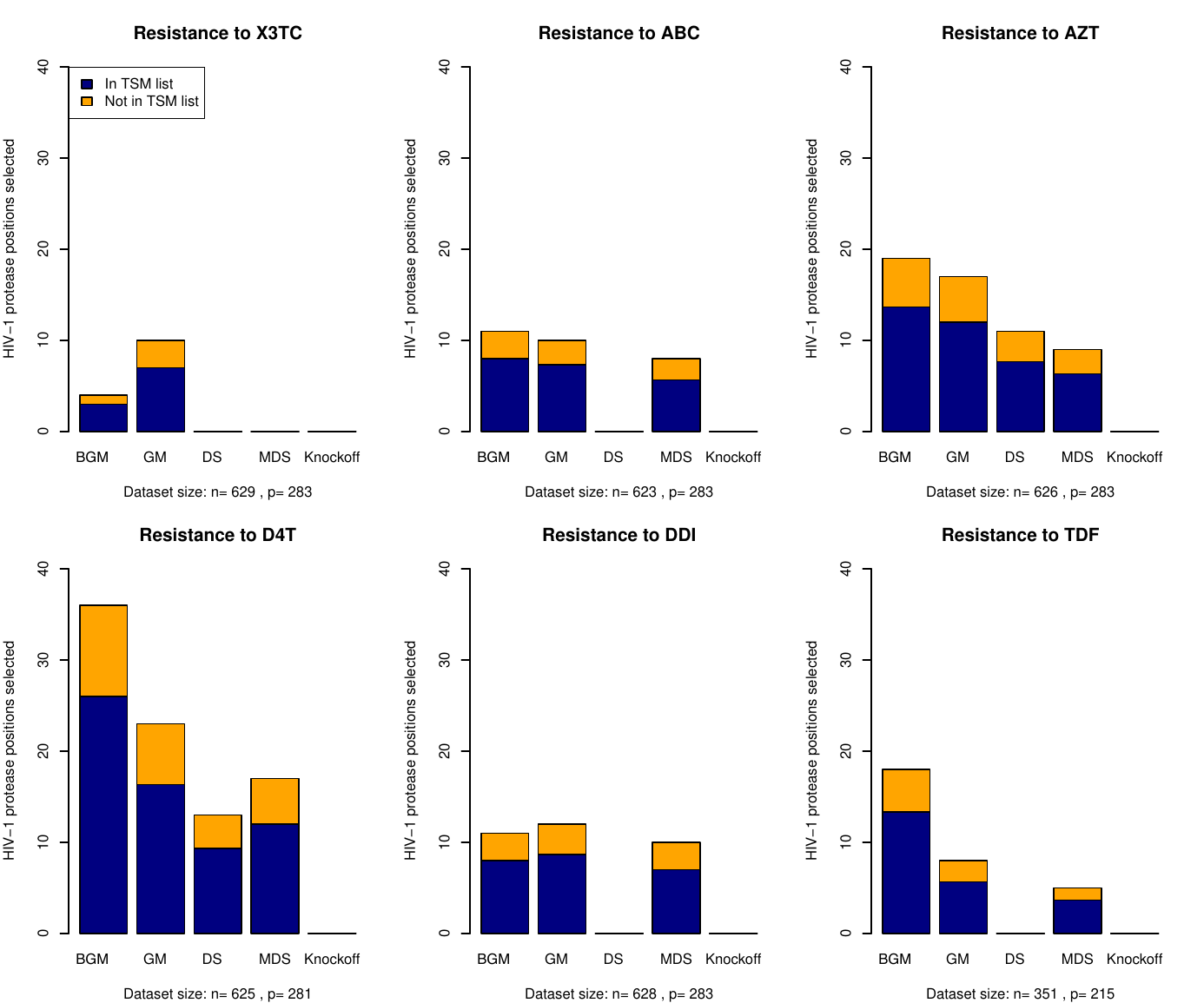}
\caption{Same as Figure~\ref{PI}, but for the six NRTI drugs.}
\label{NRTI}
\end{figure}

\begin{figure}[H]
\centering
\includegraphics[width=12cm,height=4.6cm]{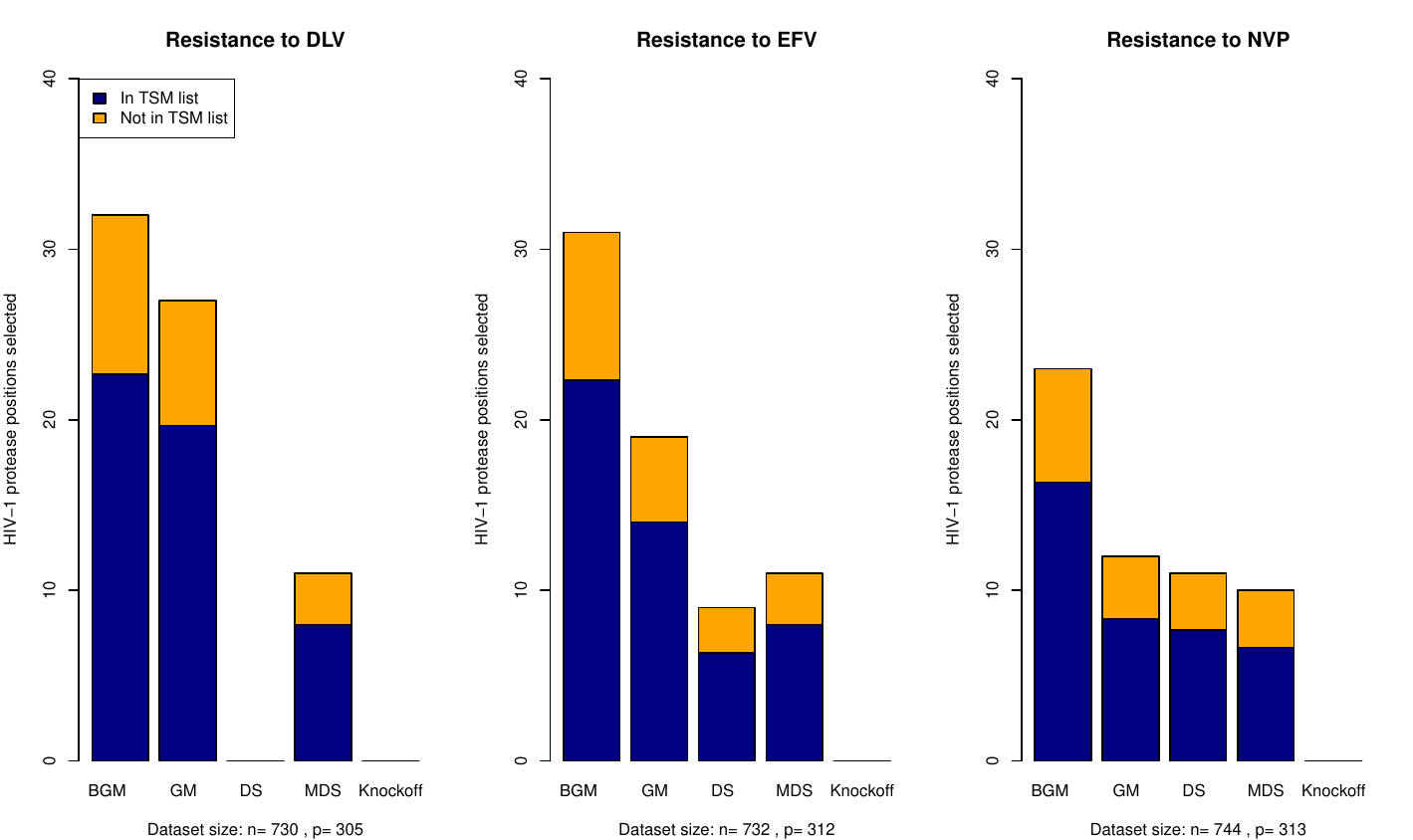}
\caption{Same as Figure~\ref{PI}, but for the three NNRTI drugs.}
\label{NNRTI}
\end{figure}

The response vector $\bm Y$ records the log-fold increase in lab-tested drug resistance. The design matrix $\bm X$ is binary, with entries $\bm X_{ij} \in \{0,1\}$, indicating the presence or absence of mutation $j$ in the $i$-th sample. We analyze each drug separately and the task is to select relevant mutations for each drug inhibitor. For each drug analysis, we process the data as follows: First, we remove patients with missing drug resistance information. Second, we retain only those mutations that appear more than three times across all patients. Consequently, the final analyzed sample size $n$ and the number of mutations $p$ vary across drugs and are slightly smaller than those reported in Table~\ref{tab:HIV-data}. However, these values are still in the hundreds, with the ratio $n/p$ ranging from 1.5 to 4 (see Figures \ref{PI}-\ref{NNRTI} for case-specific details). We assume an additive linear model between the response and features, with no interactions. The designated FDR control level is set to $q = 0.1$.

\begin{table}[H]
\centering
\caption{Description of HIV-1 datasets sizes.}
\label{tab:HIV-data}
\footnotesize
\vspace{0.2cm}
\begin{tabular}{cccc}
\hline
\multicolumn{1}{l}{Drug class}&  
\multicolumn{1}{l}{Number of drugs} & \multicolumn{1}{l}{Sample size} & 
\multicolumn{1}{l}{\vspace{0.2cm}{ Number of mutations appearing $\ge 3$ times}}\\
\hline
PI &7 &846 &209 \\
NRTI &6 &634 &287 \\
NNRTI &3 &745 &319\\
\hline
\end{tabular}
\end{table}

Following \cite{Barber2015}, we evaluate the performance of each method by assessing how many of the identified mutations overlap with the treatment-selected mutation (TSM) panels established by \cite{rhee2005hiv}. For each drug class (PIs, NRTIs, NNRTIs), the TSM panels comprise mutations that occur at significantly higher frequencies in virus samples from individuals treated with the specific drug class, compared to treatment-na\"{i}ve individuals. As such, these panels serve as a reliable approximation of the ground truth, allowing us to assess the performance of each method.

The selection results for PIs, NRTIs, and NNRTIs, are displayed in Figures \ref{PI}-\ref{NNRTI}, respectively.
For PIs, we find that the BGM method performs best for drugs APV, ATV, IDV, and RTV, and is comparable to the GM method for LPV and NFV.
In addition, the BGM method consistently outperforms the DS, MDS, and Knockoff methods across all seven drugs.
For NRTIs, Figure \ref{NRTI} shows that the BGM method performs best for drugs ABC, AZT, D4T, and TDF, while being comparable to the GM and MDS methods for drug DDI. Furthermore, the BGM method demonstrates superiority over the DS, MDS, and Knockoff methods for almost all six drugs.
Figure \ref{NNRTI} presents the corresponding results for NNRTIs, where the BGM method demonstrates the best performance across all three drugs.
In addition, we observe that the DS and Knockoff methods lack power, failing to select any mutations for the majority of drugs.

In summary, despite some variability in results across different drugs, the self-guiding BGM method effectively selects variables corresponding to true effects, demonstrating better overall agreement with the independently curated TSM lists than competing methods.

\subsection{Human breast cancer scRNA-seq}

This dataset contains gene expression profiles  from 400 T47D A12 human breast cancer cells treated with 100 nM synthetic glucocorticoid dexamethasone (Dex) at five time points (1, 2, 4, 8, and 18 hours), which is accessible at \url{https://www.ncbi.nlm.nih.gov/geo/query/acc.cgi?acc=GSE141834}. For the treatment group, scRNA-seq experiments were conducted at each time point, yielding 2,000 samples (400 cells $\times$ 5 time points). For the control group, 400 vehicle-treated cells were profiled via an scRNA-seq experiment at the 18-hour time point. The resulting dataset includes 2,400 samples. The task is to identify genes associated with the glucocorticoid response in this cell line. 

The response vector 
$\bm Y$ is binary: $\bm Y_i=1$ indicates that sample $i$ belongs to the treatment group, while $\bm Y_i=0$ denotes that sample 
$i$ is part of the control group. The design matrix $\bm X$ captures 32,049 gene expression values, where $\bm X_{ij}$ represents the expression level of the $j$-th gene in the $i$-th sample.
To reduce dimensionality, we adopt the data processing approach in \cite{hoffman2020single}. First, we filter out genes detected in fewer than 10\% of cells. Then, we select the 500 genes with the highest expression variance. This yields a final analyzed dataset with a sample size of $n=2,400$ and a feature count of $p=500$, resulting in a ratio of $n/p=4.8$. We model the relationship between the response and features using the logistic regression.

To assess the performance of each method, we use Table 1 from \cite{Dai2023} as a reference, which lists 25 genes with confirmed interactions with the glucocorticoid receptor (GR), supported by existing literature. We denote these 25 genes as the gene set $\mathcal{S}$. For each method, we compute the proportion of selected genes that are not in $\mathcal{S}$ relative to the total number of selected genes. This ratio, referred to as the approximated false selection ratio (AFSR), is calculated as: 
\begin{equation}\label{ASFR}
    \text{AFSR}=\frac{\#\{j:j\in\widehat{\mathcal{H}}_1\text{ and } j\not\in\mathcal{S}\}}{\#\{j: j\in\widehat{\mathcal{H}}_1\}}.
\end{equation}
It is sensible that a lower AFSR indicates better FDR control. In Table \ref{tab:scRNAseq-data}, we present the gene set $\mathcal{S}$ and report the selection results for each method.

Under the FDR level $q=0.1$, Table \ref{tab:scRNAseq-data} shows that the proposed BGM method outperforms the Knockoff procedure by selecting more true genes while achieving lower AFSR values. Although the DS and MDS methods identify more true features than the BGM, their AFSR values are approximately $5/16=0.3125$ and $6/26\approx 0.2308$, respectively, both exceeding the target level $q=0.1$, indicating poor FDR control. In contrast, the BGM method achieves an exact AFSR value of 0, demonstrating superior FDR control. To further evaluate performance, we relax the FDR control level to $q=0.2$ and compare the BGM and MDS methods. As shown in Table \ref{tab:scRNAseq-data}, at $q=0.2$, the BGM successfully identifies 19 of the 25 genes in the gene set $\mathcal{S}$, comparable to the MDS, which selects 23 such genes. However, the MDS yields a higher AFSR of $12/35 \approx 0.3429$,  significantly exceeding the target level. In contrast, the BGM maintains an AFSR of $4/23 \approx 0.1739$, remaining below the threshold. 
Therefore, we conclude that the proposed BGM method achieves better FDR control than the MDS in variable selection for this dataset.

From Table \ref{tab:scRNAseq-data}, we find that 21 selected genes are excluded in the gene set $\mathcal{S}$.
To verify the accuracy of $\mathcal{S}$, we use Grok 3 to verify whether these 21 genes indeed lack supporting literature. Consequently, we identify 6 selected genes that may interact with the glucocorticoid receptor through secondary effects: ATP5B, MT-ND1, NDUFA4, and COX6A1 \citep{du2009dynamic}; PSMB4 \citep{combaret2004glucocorticoids}; and YWHAZ \citep{aitken200614}.
Accordingly, we incorporate these 6 genes into an expanded gene set, denoted as $\mathcal{S}^+$, and define the adjusted approximated false selection ratio ($\text{AFSR}^+$) as follows:
\begin{equation}\label{ASFR+} \text{AFSR}^+=\frac{\#\{j:j\in\widehat{\mathcal{H}}_1\text{ and } j\not\in\mathcal{S}\cup\mathcal{S}^+\}}{\#\{j: j\in\widehat{\mathcal{H}}_1\}}. \end{equation}

\begin{table}[H]
\centering
\caption{Selection results in scRNA-seq data. Genes in {\it italics} indicate those excluded from the gene set $\mathcal{S}$, while genes in {\it\textbf{italics}} indicate those excluded from both $\mathcal{S}$ and the expanded gene set $\mathcal{S}^+$. The AFSR is computed as: $\frac{\#\{j: j\in\widehat{\mathcal{H}}_1\text{ and } j\not\in\mathcal{S}\}}{\#\{j: j\in\widehat{\mathcal{H}}_1\}}$ as given in Equation \eqref{ASFR}, and $\text{AFSR}^+$ is calculated as: $\frac{\#\{j: j\in\widehat{\mathcal{H}}_1\text{ and } j\not\in\mathcal{S}\cup\mathcal{S}^+\}}{\#\{j: j\in\widehat{\mathcal{H}}_1\}}$ as given in Equation \eqref{ASFR+}.}
\label{tab:scRNAseq-data}
\tiny
\vspace{0.2cm}
\begin{tabular}{cccccc}
\hline
\multicolumn{1}{c}{Gene set $\mathcal{S}$:} & 
\multicolumn{5}{c}{\makecell{ SERPINA6, FKBP5, NFKBIA, RPL10, SEMA3C, HSPB1, RBBP7, C1QBP, \\EIF4EBP1, S100A11, NUPR1, MSX2, HSPA8, HSPA1A, EEF1A1, RBM24, BCL6, \\ATF4, IGFBP4, YWHAQ, DDIT4, IRX2, GATA3-AS1, TACSTD2, DSCAM-AS1}} \\
\hline
\multicolumn{1}{c}{Expanded gene set $\mathcal{S^+}$:} & 
\multicolumn{5}{c}{ATP5B, PSMB4, YWHAZ, NDUFA4, MT-ND1, COX6A1} \\
\hline 
\multicolumn{1}{c}{FDR level} & 
\multicolumn{1}{c}{Method} & 
\multicolumn{1}{c}{Selected genes} & 
\multicolumn{1}{c}{\makecell{Number of \\selected genes}} &
\multicolumn{1}{c}{AFSR} &
\multicolumn{1}{c}{$\text{AFSR}^+$} \\
\hline
\multirow{4}[35]{*}{$q=0.1$ }
& BGM &\makecell{SERPINA6, FKBP5, RPL10, ATF4, \\DSCAM-AS1, S100A11, NFKBIA} & 7 & 0 &0\\
\cline{2-6}
& DS &\makecell{SERPINA6, FKBP5, NFKBIA, ATF4, RPL10, \\SEMA3C, DSCAM-AS1, S100A11, RBBP7, IGFBP4, \\EIF4EBP1, {\it \textbf{RPL28, PIP, RPL30, RPL19,}} {\it ATP5B}} &16 & 5/16 & 4/16\\
\cline{2-6}
& MDS &\makecell{SERPINA6, FKBP5, NFKBIA, RPL10, SEMA3C, \\HSPB1, S100A11, ATF4, DSCAM-AS1, RBBP7, \\EIF4EBP1, IGFBP4, YWHAQ, DDIT4, IRX2, \\GATA3-AS1, RBM24, TACSTD2, EEF1A1, BCL6, \\{\it \textbf{RPL41, SNRPB, SF3B5, RPLP0P6, SNHG19, UQCR10}}} &26 & 6/26 & 6/26\\
\cline{2-6}
& Knockoff & \makecell{SERPINA6, NFKBIA, {\it \textbf{RPL28, VAMP8}, PSMB4}}& 5 &3/5 & 2/5 \\
\hline
\multirow{2}[25]{*}{$q=0.2$ }
& BGM &\makecell{SERPINA6, FKBP5, RPL10, ATF4, HSPB1, \\DSCAM-AS1, S100A11, NFKBIA, SEMA3C, \\RBBP7, EIF4EBP1, YWHAQ, DDIT4, IRX2, \\RBM24, TACSTD2, EEF1A1, BCL6, C1QBP, \\{\it\textbf{RPS18}, YWHAZ, NDUFA4,  ATP5B}} &  23 &4/23 &1/23\\
\cline{2-6}
& MDS &\makecell{SERPINA6, FKBP5, DDIT4, IRX2, NFKBIA, \\HSPB1, S100A11, ATF4, GATA3-AS1, RBBP7, \\EIF4EBP1, IGFBP4, YWHAQ, RPL10, SEMA3C, \\DSCAM-AS1, RBM24, TACSTD2, EEF1A1, BCL6, \\NUPR1, MSX2, C1QBP, {\it \textbf{C1orf43, RPL41, SNRPB,}} \\{\it \textbf{RPLP0P6, SNHG19, PPDPF, TMEM141,}} \\{\it \textbf{SF3B5, UQCR10,} MT-ND1, COX6A1, PSMB4}} & 35 &12/35 &9/35\\
\hline
\end{tabular}
\end{table}


Similar to AFSR, a lower $\text{AFSR}^+$ value indicates better FDR control.
From Table \ref{tab:scRNAseq-data}, we observe that under the FDR level $q=0.1$, the $\text{AFSR}^+$ results are consistent with those based on AFSR. At the FDR level $q=0.2$, both the BGM and MDS methods identify 3 additional genes included in the expanded gene set $\mathcal{S}^+$. While the $\text{AFSR}^+$ of the MDS method decreases to $9/35\approx 0.2571$ compared to its AFSR value ($\approx 0.3429$), it still exceeds the set threshold of $q=0.2$. In contrast, the BGM method maintains an $\text{AFSR}^+$ of $1/23\approx 0.0435$, which is significantly lower than its AFSR value ($\approx 0.1739$) and remains well below the target level. Therefore, when evaluated using the expanded gene set, the proposed BGM method demonstrates superior FDR control compared to other methods, even surpassing its performance based on the AFSR assessment.

Overall, the proposed BGM method performs best by selecting more genes with supporting evidence while effectively controlling FDR. Note that the empirical results for the DS and MDS methods shown here differ slightly from those in \cite{Dai2023}, expected due to the random sampling process and non-deterministic nature of these algorithms.

\section{Concluding remarks}

In this paper, we develop an easily implementable procedure, namely self-guiding BGM, for FDR control in high-dimensional variable selection problems. The proposed framework relaxes the constraints on model settings and data distributions, while also avoiding substantial power loss. The key innovation of our method lies in its ability to operate without requiring the test statistic to be symmetric around zero under the null hypothesis. This unique feature directly addresses a critical limitation of many existing methods. Through theoretical and empirical analyses, we demonstrate that our method can asymptotically control the FDR while remaining powerful against alternatives under mild conditions.
{For future work, an interesting direction is to extend our BGM method under a model-free setting.}


\scsection{Supplementary materials}
The supplement contains the proofs of theorems and propositions.

\bibliographystyle{chicago}
\bibliography{Ref}

\end{document}